\documentclass[journal]{IEEEtran}

\ifCLASSINFOpdf
\else
   \usepackage[dvips]{graphicx}
\fi
\usepackage{url}
\usepackage{amsmath,graphicx}
\usepackage{float}
\usepackage{booktabs}
\usepackage{multirow}
\usepackage{tabularx}
\usepackage{float}
\usepackage{pifont}
\usepackage{amssymb}
\usepackage{lipsum} 
\usepackage{enumitem} 
\usepackage{bbding}
\usepackage{threeparttable} 
\usepackage{bm}

\usepackage[breaklinks]{hyperref}
\usepackage{CJKutf8}

\usepackage{times}
\usepackage{latexsym}

\usepackage[T1]{fontenc}

\usepackage[utf8]{inputenc}

\usepackage{microtype}

\usepackage{inconsolata}

\usepackage{graphicx}

\begin{document}

\title{Echotune: A Modular Extractor Leveraging the Variable-Length Nature of Speech in ASR Tasks}

\author{
    Sizhou Chen, 
    Songyang Gao,
    and Sen Fang
\thanks{Manuscript received January 31, 2024.}
\thanks{Sizhou Chen is with the Blockchain Industry School, Chengdu University of Information Technology, Chengdu 610225, China (e-mail: szchen1005@gmail.com).}
\thanks{Songyang Gao is with the School of Computer Science and Technology, Fudan University, Shanghai 200433, China (e-mail: gaosy21@m.fudan.edu.cn).}
\thanks{Sen Fang is with the Department of Computer Science, Victoria University, PO Box 14428, Melbourne, VIC 8001, Australia (e-mail: sen.fang@live.vu.edu.au).}}

\markboth{Journal of \LaTeX\ Class Files, Vol. 14, No. 8, August 2015}
{Shell \MakeLowercase{\textit{et al.}}: Bare Demo of IEEEtran.cls for IEEE Journals}
\maketitle

\begin{abstract}
The Transformer architecture, pivotal in Automatic Speech Recognition (ASR), traditionally uses fixed-length attention windows, limiting its effectiveness with varied speech sample durations and complexities. This often leads to data over-smoothing and misses long-term connections in speech. To overcome this, we introduce Echo Multi-Scale Attention (Echo-MSA), a module with a variable-length attention mechanism adaptable to different speech complexities and durations. It can extract speech features at multiple levels, from frames and phonemes to words and discourse, addressing the limitations of fixed-length attention. Our design uses a parallel attention structure with a dynamic gating mechanism, blending traditional attention with the output of Echo-MSA. This integration significantly improves the word error rate (WER) performance while maintaining the stability of the original model, as demonstrated by our empirical studies.

\end{abstract}


\IEEEpeerreviewmaketitle

\section{Introduction}

\IEEEPARstart{I}{n} the area of speech recognition, Transformer has gained recognition for its ability to manage long-term dependencies in automatic speech recognition (ASR) tasks~\cite{c1}. Prior studies, like HMM-DNN~\cite{c2} and HMM-GMM~\cite{c3}, typically involved numerous modules and steps, whereas end-to-end ASR systems~\cite{c4,c5,c9} employed immediate audio-to-text mapping. Nevertheless, there are restrictions to the use of Transformer in ASR~\cite{c6,c7}. The rise of multimodal information integration has increased attention towards developing self-supervised models.  

In the recent past, speech recognition has witnessed progress due to self-supervised pretraining models. Notably, Wav2vec 2.0~\cite{c29} uses exclusively unlabeled data for pre-training, thus efficiently learning semantically aligned speech sequence representations. Other models, such as HuBERT~\cite{c16} and Data2Vec~\cite{c11}, aim to predict hidden speech representations and accurately map speech to semantic space through a speech segment prediction task, respectively. Nevertheless, speech signals inherently contain multiple attributes with interconnected multimodal information. Unfortunately, existing modeling techniques still have limitations in capturing this information, highlighting the need for continued exploration and refinement of these techniques. 

A model's complete comprehension of speech is tied to its treatment of short and long signals. Liu~\cite{c31} posit that implementing the Attention mechanism may result in over-smoothing, which can blur crucial information amidst speech segment length variations. Wang~\cite{c32} proposes that a self-attention window of fixed length may overlook significant long-term connections. All of these studies indicate the necessity of speech recognition models with the ability to handle inputs of varying lengths.

We believe that crafting adaptable models to address the variable length traits of speech is fundamentally essential to solving this issue. This insight is rooted in Echo Multi-Scale Attention (Echo-MSA), depicted in Fig.\ref{fig:f2}. It uses dynamic attention for speech sequences of varying lengths, extracting speech features at different details and enhancing its modeling of variable-length speech features. Experiments show that Echo-MSA boosts the stability and accuracy of speech recognition.

Our main contributions are threefold:

\begin{enumerate}[label=(\arabic*),leftmargin=*,itemsep=0.1em,parsep=0.1em] 
    \item We introduce Echo-MSA, a modular extractor designed for speech recognition that enhances the accuracy of representing speech information.
    \item We enable seamless integration of Echo-MSA with underlying models by combining attentional parallelism techniques and hybrid loss. 
    \item In the Librispeech dataset, Echo-MSA is integrated into the backbone network. We conduct thorough experimental analyses to verify the effectiveness of Echo-MSA and the training process.
\end{enumerate}

The following section provides an overview of the data2vec backbone model and the pre-existing Speech-Transformer. Section 3 presents the proposed module along with its training methodology. We detail the experimental setup in Section 4 and analyze the findings in Section 5. Lastly, the paper concludes in Section 6.

\begin{figure}[h] 
  \centering 
  \includegraphics[width=0.18\textwidth]{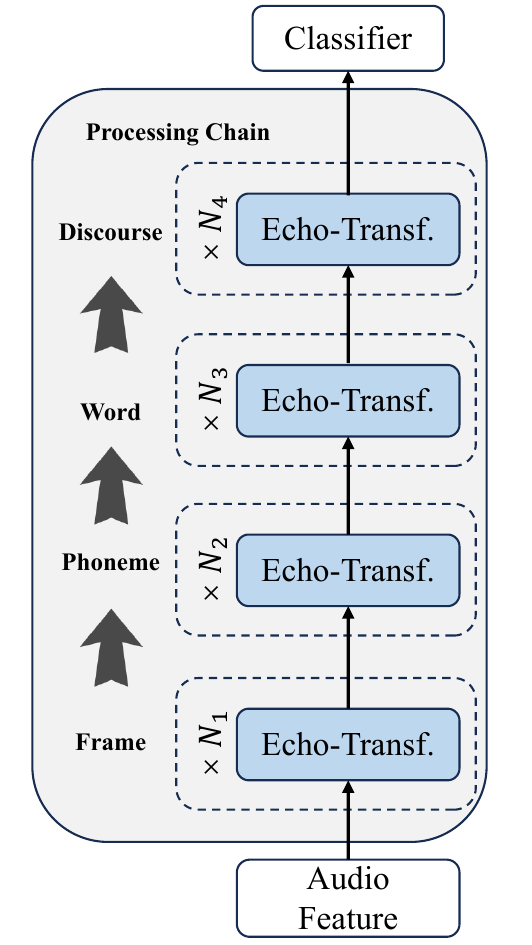} 
    \vspace{-10pt}
  \caption{Hierarchical Echo-Transformer Training Framework with Multi-Stage Processing.}
  \label{fig:f1}
  \vspace{-1.5em}
\end{figure}

\vspace{-1em}
\section{RELATED WORK}
\label{sec:format}

Data2vec, a multimodal framework, draws inspiration from Wav2vec~\cite{c8} and HuBERT~\cite{c16}. It employs contrast learning~\cite{c11} for self-supervision and extracts features from speech, images, and text label-free. Unlike Wav2vec or HuBERT, focused solely on speech, Data2vec learns to correlate multimodal data and share insights. It excels over other unsupervised methods, such as Skip-thought~\cite{c17}, in multimodal learning. For tasks like speech recognition, specialized models might be more effective.  

In speech attention, advancements comprise Zhang~\cite{c19}'s deployment of deep networks for enhanced ASR. Dong's 2D-Attention~\cite{c20} sharpens Speech-Transformer's focus, and Ramabhadran~\cite{c21} added multiple softmaxes to amplify attention in Transformers. Yet, these methods overlook the variable length character of speech. We outline our specific enhancements in the subsequent section. 

\vspace{-1em}

\section{METHODOLOGY}
\label{sec:pagestyle}

\subsection{Model Architecture}
\label{ssec:subhead}

The training framework, depicted in Figure \ref{fig:f1}, includes Echo-Transformer blocks with four Echo-MSA attention mechanisms, detailed further in Figure \ref{fig:f2}. The DualFocusGate integrates Echo-MSA with standard MSA, allowing flexible switching between Echo-MSA and Self-Attention, enhancing speech data analysis by capturing statistical features.

In our training methodology, we employ a compound loss function $\mathcal{L}_{\mathrm{E-ctc}}$, which amalgamates class-weighted Connectionist Temporal Classification (CTC) with Focal Loss~\cite{c33}. This integration is pivotal for mitigating class imbalance in Automatic Speech Recognition (ASR) tasks. Focal Loss plays an integral role in modulating the loss function, diminishing the emphasis on prevalent and easily classifiable instances while augmenting the focus on infrequent and intricate cases.

The compound loss function is represented as:

\resizebox{0.96\hsize}{!}{$
\begin{array}{ll}
\mathcal{L}_{\mathrm{E-ctc}}=\lambda \mathcal{L}_{\mathrm{W-ctc}}+(1-\lambda) F(x) & \text{(1)} \\
\hspace{0.25cm} \mathcal{L}_{\mathrm{W-ctc}}=\frac{1}{N} \sum_{i=1}^{N}\left(L_{\mathrm{CTC}, i} \times w_{i}\right) & \text{(2)} \\
\hspace{0.25cm} \mathcal{L}_{\mathrm{CTC}}=-\log \left(\sum_{\pi \in \operatorname{AllAlignments}(y)} P(\pi \mid \epsilon)\right) & \text{(3)} \\
\hspace{0.25cm} F(x)=\alpha \sum_{i}\left(1-e^{-x_{i}}\right)^{\gamma} x_{i} & \text{(4)}
\end{array}
$}

\noindent $\lambda$ serves as a weight adjuster for CTC and Focal Loss, initially set at 0.5. $\mathcal{L}_{\mathrm{W-ctc}}$ represents the weighted CTC loss, while $F(x)$ tackles category imbalance through Focal Loss. $\alpha$, set at 0.25, balances the weights of the samples, and $\gamma$, valued at 2, reduces the loss for easily classifiable samples. $N$ denotes the count of samples in the batch, with $w_i$ representing the weight of the $i$-th sample. In $\mathcal{L}_{\mathrm{CTC}}$, $x$ denotes the model's log probability output for a specific phoneme or word, which is used to modulate the loss contribution, and $y$ is the target label, with $\mathrm{AllAlignments(y)}$ indicating the possible alignments of $y$. The probability of a specific alignment $\pi$ given $x$ is denoted by $P(\pi \mid \epsilon)$. $F(x)$ performs an operation on each element $x_i$ of vector $x$, contributing to the total loss.

\vspace{-0.5em}

\subsection{Echo-MSA}
\label{ssec:subhead}


\begin{figure*}[t!] 
  \centering
  \includegraphics[width=0.65\textwidth]{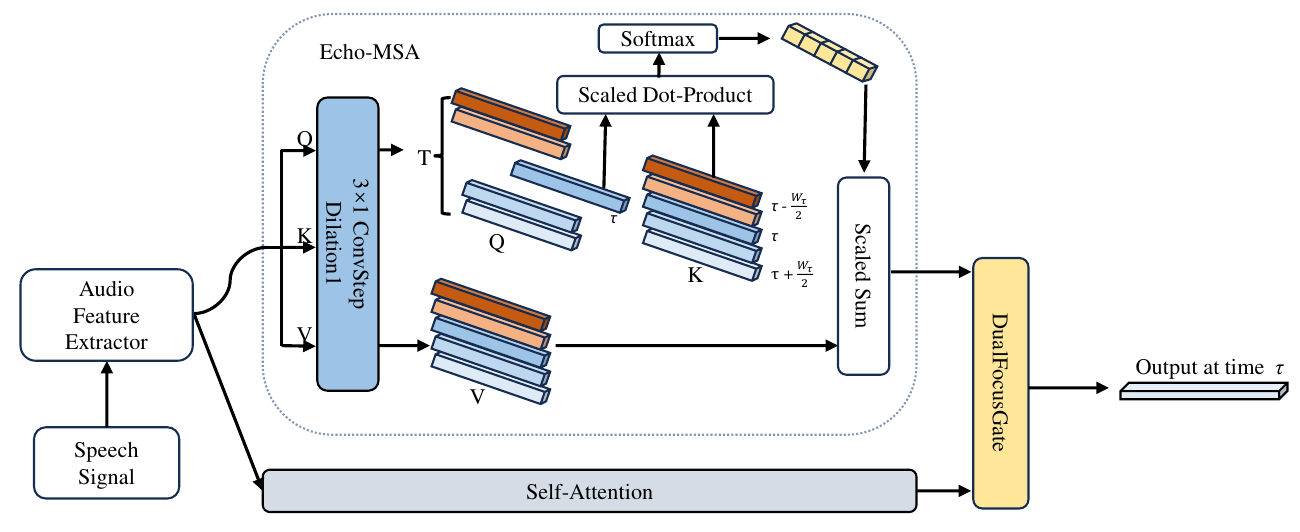}
  \vspace{-5pt}
  \caption{Embedding Echo-MSA with Variable-Length Multi-Scale Attention into Pretrained Models Assisted by Dual Focus Gate at Time Step $\tau$, where $W_{\phi}$ Represents Customizable Variable Length.}
  \label{fig:f2}
  \vspace{-1em}
\end{figure*}

As depicted in Fig.\ref{fig:f2}, Echo-MSA processes data via a depth-separable convolutional layer, expanding the receptive field to capture global speech signal details. It uses $W_{\phi}$ for fine-grained extraction, where applying window $W_{\phi}$ limits full-attention computation to few neighboring tokens, reducing computational load. Echo-MSA also allows personalized learning by varying $W_{\phi}$ values in different Transformer stages, understanding interactions between frames, phonemes, and words. The complete Echo-MSA output is calculated by:

\noindent \textbf{Step I:} Based on the current stage level, we apply a specific window \( W_{\phi} \). Further details regarding the selection and values of \( W_{\phi} \) are elaborated in Section 4.2.\\
\noindent \textbf{Step II:} The Key (\( K \)), Value (\( V \)), and Query (\( Q \)) at the current time step are fed into a depthwise separable convolution to reduce both the model's parameter count and computational complexity. \\ 
\noindent \textbf{Step III:} We select tokens from the \( \tau - \frac{W_{\phi}}{2} \)th to the \( \tau + \frac{W_{\phi}}{2} \)th positions in \( K \). Each of these tokens performs a scaled dot-product with the \( \tau \)-th token in \( Q \) to generate scores. All scores are concatenated and scaled via a Softmax function to produce the attention weights. \\
\noindent \textbf{Step IV:} Tokens from the \( \tau - \frac{W_{\phi}}{2} \)th to the \( \tau + \frac{W_{\phi}}{2} \)th positions in \( V \) are retrieved, and the sum of each token multiplied by its weight is computed. The result serves as the output for the \( \tau \)-th token in Echo-MSA. \\ 
\noindent \textbf{Step V:} Return to Step III and continue until \( \tau \) iterates from 1 to \( T \), where \( T \) denotes the length of the input sequence in Echo-MSA. In the context of ASR, \( T \) typically represents the number of frames or processed speech segments in the speech signal.

\vspace{-1em}
\subsection{Dual Focus Gate}
\label{ssec:subhead}

In the Echo-Transformer framework, integrating new modules with pre-trained weights is essential. This is achieved using a feed-forward network. With input matrix \( \mathbf{X} \) and attention mask \( \mathbf{M} \), the Multi-Scale Attention (MSA) produces outputs \( \mathbf{O_{1}} \) and \( \mathbf{A_{1}} \), while Echo-MSA outputs \( \mathbf{O_{2}} \).


\begin{align}
\mathbf{O}_{1}, \mathbf{A}_{1} &= \text{MSA}(\mathbf{X}, \mathbf{M}) \tag{5} \\ 
\mathbf{O}_{2} &= \text{Echo-MSA}(\mathbf{X}) \tag{6}
\end{align}

The Dual Focus Gate, with ReLU and Sigmoid activations, uses two layers. It computes intermediate \( \mathbf{H} \) from \( \mathbf{X} \) and \( \mathbf{b}_{1} \), and \( \mathbf{Z} \) from \( \mathbf{H} \) and \( \mathbf{b}_{2} \), deriving \( \mathbf{G} \).


\begin{align}
\mathbf{H}&=\operatorname{ReLU}\left(\mathbf{W}_{1} \mathbf{X}+\mathbf{b}_{1}\right) \tag{7} \\
\mathbf{Z}&=\mathbf{W}_{2} \mathbf{H}+\mathbf{b}_{2} \tag{8} \\
\mathbf{G}&=\sigma(\mathbf{Z}) \tag{9}
\end{align}

The final output \( \mathbf{O}_{out} \) combines \( \mathbf{O}_{1} \) and \( \mathbf{O}_{2} \), weighted by \( \mathbf{G} \), balancing the attention mechanisms' outputs.


\begin{equation}
\mathbf{O}_{out} = \mathbf{G} \odot \mathbf{O}_{1} + (\mathbf{1}-\mathbf{G}) \odot \mathbf{O}_{2} \tag{10}
\end{equation}

The modulation results in \( \mathbf{O}_{out} \) being a balanced mix of attention outputs, preserving the original input's integrity and integrating Echo-Transformer's new insights.

\vspace{-0.5em}
\section{EXPERIMENTAL SETUP}
\label{sec:typestyle}

\subsection{Dataset}
\label{ssec:subhead}

We conducted comprehensive experiments on the LibriSpeech corpus~\cite{c23}, including both the 100-hour "clean" subset and the 960-hour full dataset. For evaluation, we used four test sets: dev-clean, dev-other, test-clean, and test-other, ensuring a thorough investigation. We also employed 60,000 hours of unlabeled Libri-light corpus~\cite{c24} data as an auxiliary resource. 

\vspace{-1em}

\subsection{Model architecture and training recipe}
\label{ssec:subhead}
\begin{table*}[t]  
    \centering
    \vspace{-10pt}
    \caption{Performance Metrics of ASR (Automatic Speech Recognition) on LibriSpeech Development and Test Sets Using a 100-Hour clean Training Subset.} 
    \label{table:t1}
    \renewcommand{\arraystretch}{0.8}  
    \small
    \resizebox{0.7\textwidth}{!}{
    \begin{tabular}{ccccccc}
        \toprule
        \multirow{2}{*}{Model} & \multirow{2}{*}{Unlabeled} & \multirow{2}{*}{LM} & \multicolumn{2}{c}{dev} & \multicolumn{2}{c}{test} \\
        \cmidrule(lr){4-5}  
        \cmidrule(lr){6-7}  
        &  data &  & clean & other & clean & other \\
        \midrule
        Noisy student~\cite{c26} & LS-860 & LSTM & 3.9 & 8.8 & 4.2 & 8.6 \\
        IPL~\cite{c27} & LL-60K & 4-gram+Transf. & 3.2 & 6.1 & 3.7 & 7.1 \\
        SlimIPL~\cite{c28} & LS-860 & 4-gram+Transf. & 2.2 & 4.6 & 2.7 & 5.2 \\
        \midrule
        DiscreteBERT~\cite{c25} & LS-960 & 4-gram & 4 & 10.9 & 4.5 & 12.1 \\
        wav2vec 2.0(Base)~\cite{c29} & LS-960 & 4-gram & 2.7 & 7.9 & 3.4 & 8.6 \\
        Hubert(Base)~\cite{c16} & LS-960 & 4-gram & 2.7 & 7.8 & 3.4 & 8.1 \\
        data2vec(Base)~\cite{c11} & LS-960 & 4-gram & 2.6 & 7 & 2.8 & 7 \\
        Our Model(Base) & LS-960 & 4-gram & \textbf{2.4} & \textbf{6.6} & \textbf{2.5} & \textbf{6.6} \\      
        \midrule
        data2vec(Large)~\cite{c11} & LL-60K & 4-gram & 1.9 & 3.9 & 1.9 & 4.1 \\  
        Our Model(Large) & LL-60K & 4-gram & \textbf{1.7} & \textbf{3.9} & \textbf{1.7} & \textbf{3.7} \\
        \bottomrule
    \end{tabular}
    }
\end{table*}

Experiments used the Huggingface Transformers library~\cite{c22}. Baseline models, data2vec (Base) and (Large), are 'data2vec-audio-base' and '-large' on Huggingface. Analyses with Baseline models employed these specific pre-trained versions.

Our experiments involved two Echo-Transformer models: a 12-layer Base model (Echo-S configuration, $N_1$ $\sim$ $N_4$ $=$ $\{2, 2, 4, 4\}$, $W_{\phi}$ $=$ $\{4, 16, 64, 256\}$) and a 24-layer Large model (Echo-B configuration, $N_1$ $\sim$ $N_4$ $=$ $\{4, 4, 8, 8\}$, $W_{\phi}$ $=$ $\{4, 16, 64, 256\}$). Both models processed 16 kHz audio through a function encoder as detailed in~\cite{c29}, outputting at 50 Hz with a 20-millisecond sample interval and normalizing input waveforms. This demonstrates the Echo-Transformer's scalability.



In ASR model training, we applied a stage-based learning rate strategy with three rates (6e-5, 6e-6, 6e-7) at different stages. These rates, combined with cosine annealing scheduling and a weight decay of 0.0005, enhanced model regularization and training efficiency.

\vspace{-1em}

\section{RESULTS AND ANALYSIS}
\label{sec:subhead}


\subsection{Results on the 100-hour train data}
\label{ssec:subhead}


Table \ref{table:t1} demonstrates the efficacy of our ASR model, "Our Model," post fine-tuning with the LibriSpeech 100/960 hour datasets. This model integrates the Echo-MSA module into the data2vec framework~\cite{c11} and is compared against leading self-supervised learning methods, including DiscreteBERT~\cite{c25}, Noisy Student~\cite{c26}, IPL~\cite{c27}, and HuBERT~\cite{c16}, with a focus on our Baseline model.

For the Base configuration, Our Model(Base) outperforms data2vec(Base), attaining a WER of 2.4 (clean) and 6.6 (other) against 2.6 and 7, respectively, yielding WERRs of 7.7\% (clean) and 5.7\% (other). These results emphasize Our Model's enhanced capability under complex acoustic scenarios.

In the Large model category, Our Model(Large) surpasses data2vec(Large) in both test sets. It achieves a WER of 1.7 (clean) and 3.7 (other), compared to 1.9 and 4.1 by data2vec(Large), corresponding to WERRs of 10.5\% (clean) and 9.8\% (other).


\vspace{-1em}

\subsection{Ablation between different components.}
\label{ssec:subhead}


In this section, an ablation study is presented to assess the influence of various components on the performance of Our Model(Base). The study concentrates on evaluating the differential impact of two loss functions, $\mathcal{L}_{\mathrm{CTC}}$ and $\mathcal{L}_{\mathrm{E-CTC}}$, and the incorporation of Echo-MSA and Dual Focus Gate, on the Word Error Rate (WER) for datasets comprising 1-hour and 100-hour labeled data.

\begin{table}[h!]
    \centering
    \vspace{-10pt}
    \caption{Ablation Study on Base Version of Our Model: Impact of $\mathcal{L}_{\mathrm{CTC}}$, $\mathcal{L}_{\mathrm{E-CTC}}$, Echo-MSA, and Dual Focus Gate on Word Error Rate (WER) for 1h and 100h Labeled Data}
    \label{table:t3}
    \small
    \begin{tabular}{c|cccc}
        \hline
        \textbf{Component} & \multicolumn{4}{c}{\textbf{Choice}} \\
        \hline
        $\mathcal{L}_{\mathrm{CTC}}$ & \Checkmark &  & \Checkmark &  \\
        $\mathcal{L}_{\mathrm{E-CTC}}$ &  & \Checkmark &  & \Checkmark \\
        Echo-MSA &  &  & \Checkmark & \Checkmark \\
        Focus Gate &  &  & \Checkmark & \Checkmark \\
        \hline
        Our Model(1h) & 9.7 & 9.6 & 9.4 & 9.3 \\
        Our Model(100h) & 7 & 6.8 & 6.7 & 6.6 \\
        \hline
    \end{tabular}
    \vspace{-1em}
\end{table}


As delineated in Table \ref{table:t3}, the ablation study examined four configurations, including standalone $\mathcal{L}_{\mathrm{CTC}}$, standalone $\mathcal{L}_{\mathrm{E-CTC}}$, and their combinations with Echo-MSA and Dual Focus Gate. The study predominantly concentrated on the augmented standard CTC loss function $\mathcal{L}_{\mathrm{E-CTC}}$ and the adaptively functioning Dual Focus Gate.

The evaluation revealed that employing $\mathcal{L}_{\mathrm{CTC}}$ alone resulted in Word Error Rates (WERs) of 9.7 for 1-hour data and 7 for 100-hour data. Utilizing $\mathcal{L}_{\mathrm{E-CTC}}$ improved WERs to 9.6 (1 hour) and 6.8 (100 hours). Notably, the combination of $\mathcal{L}_{\mathrm{E-CTC}}$ with Echo-MSA and Dual Focus Gate led to the most significant WER reductions, achieving 9.3 (1 hour) and 6.6 (100 hours).

\vspace{-1em}

\subsection{Results on the low-resource labeled data}
\label{ssec:subhead}
To comprehend the efficiency of Echo-MSA in diverse resource settings, we optimized the automatic speech recognition model using labeled data ranging from 10 minutes to 100 hours. Table \ref{table:t2} evaluates various models like HuBERT~\cite{c16}, WavLM~\cite{c30}, wav2vec 2.0~\cite{c29}, and our potent baseline~\cite{c11}.

\begin{table}[h]  
    \centering
    \caption{Word Error Rate in Librispeech Test-Other: Fine-Tuning Effects of Models Pre-Trained on Diverse Datasets (LS-960, LL-60K, MIX-94K) Using Libri-light Low-Resource Labeled Data (10 min, 1 h, 100h) and Associated Language Model (LM) Descriptions.}
    \label{table:t2}
    \setlength{\tabcolsep}{3.5pt} 
    \renewcommand{\arraystretch}{1} 
    \small
    \resizebox{\columnwidth}{!}{
    \begin{tabular}{cccccc}
        \toprule
        & \multirow{2}{*}{\begin{tabular}{c}Unlabeled \\ data\end{tabular}} & \multirow{2}{*}{LM} & \multicolumn{3}{c}{Amount of labeled data} \\
        &  & & \multicolumn{1}{l}{10m} & \multicolumn{1}{c}{\hspace{0.4cm}1h} & \multicolumn{1}{r}{100h} \\
        \midrule
        \textit{Base models} &  &  &  &  &  \\
        wav2vec 2.0~\cite{c29}  & LS-960 & 4-gram & 15.6                  & \hspace{0.4cm}11.3 & \multicolumn{1}{c}{\hspace{0.2cm}8}   \\
        HuBERT~\cite{c16}       & LS-960 & 4-gram & 15.3                  & \hspace{0.4cm}11.3 & \multicolumn{1}{c}{\hspace{0.2cm}8.1} \\
        WavLM~\cite{c30}        & LS-960 & 4-gram & \multicolumn{1}{r}{-} & \hspace{0.4cm}10.8 & \multicolumn{1}{c}{\hspace{0.2cm}7.7} \\
        data2vec~\cite{c11}     & LS-960 & 4-gram & 12.3                  & \hspace{0.4cm}9.7  & \multicolumn{1}{c}{\hspace{0.2cm}7}   \\
        \midrule
        Our Model    & LS-960  & 4-gram & \textbf{11.8}                  & \hspace{0.4cm}\textbf{9.3}  & \multicolumn{1}{c}{\hspace{0.2cm}$\bm{6.6}^{\bm{*}}$} \\
        \midrule
        \textit{Large models} &         &        &                       &      &     \\
        wav2vec 2.0~\cite{c29}  & LL-60K  & 4-gram & 10.3                  & \hspace{0.4cm}7.1  & \multicolumn{1}{c}{\hspace{0.2cm}4.6} \\
        HuBERT~\cite{c16}       & LL-60K  & 4-gram & 10.1                  & \hspace{0.4cm}6.8  & \multicolumn{1}{c}{\hspace{0.2cm}4.5} \\
        WavLM~\cite{c30}        & MIX-94K & 4-gram & \multicolumn{1}{r}{-} & \hspace{0.4cm}6.6  & \multicolumn{1}{c}{\hspace{0.2cm}4.6} \\
        data2vec~\cite{c11}     & LL-60K  & 4-gram & 9.1                   & \hspace{0.4cm}5.6  & \multicolumn{1}{c}{\hspace{0.2cm}4.1} \\
        \midrule
        Our Model    & LL-60K  & 4-gram & \textbf{8.8}                     & \hspace{0.4cm}\textbf{5.3}  & \multicolumn{1}{c}{\hspace{0.2cm}\textbf{3.7}}  \\
        \bottomrule
    \end{tabular}
    }
    \begin{tabular}{p{\linewidth}}
    \footnotesize
    Note: In this table, `*` indicates results are significant at $p < 0.01$. Significance testing was selectively conducted for the 100h data, where Our Model showed significance over Baseline ($t = -2.595$, $p = 0.0095$, $variance ±0.007$).
    \vspace{-2em}
    \end{tabular}

\end{table}

Within the base model framework, Our Model exhibits a marked enhancement in performance relative to the Baseline. Utilizing 10 minutes of labeled data, Our Model attains a WER of 11.8, signifying a 4.1\% enhancement over the Baseline's WER of 12.3. With the expansion of labeled data to 1 hour, Our Model records a WER of 9.3, surpassing the Baseline's 9.7 by 4.1\%. Notably, at the 100-hour data mark, Our Model substantially lowers the WER to 6.6, a 5.7\% improvement in comparison to the Baseline's 7.

In scenarios involving larger models that incorporate LL-60K as unlabeled data, Our Model consistently surpasses the Baseline. It records WERs of 8.8, 5.3, and 3.7 for 10 minutes, 1 hour, and 100 hours of labeled data, respectively. These figures represent advancements of 3.3\%, 5.4\%, and 9.8\% relative to the Baseline's WERs of 9.1, 5.6, and 4.1 for equivalent volumes of labeled data.

The superior performance of Our Model is consistently observed across varying data scales and experimental conditions, underlining its robustness and efficacy in diverse speech recognition environments.

\vspace{-1em}
\subsection{Results on the different kernel sizes}
\label{ssec:subhead}

\begin{figure}[t] 
  \centering
  \includegraphics[width=0.3\textwidth]{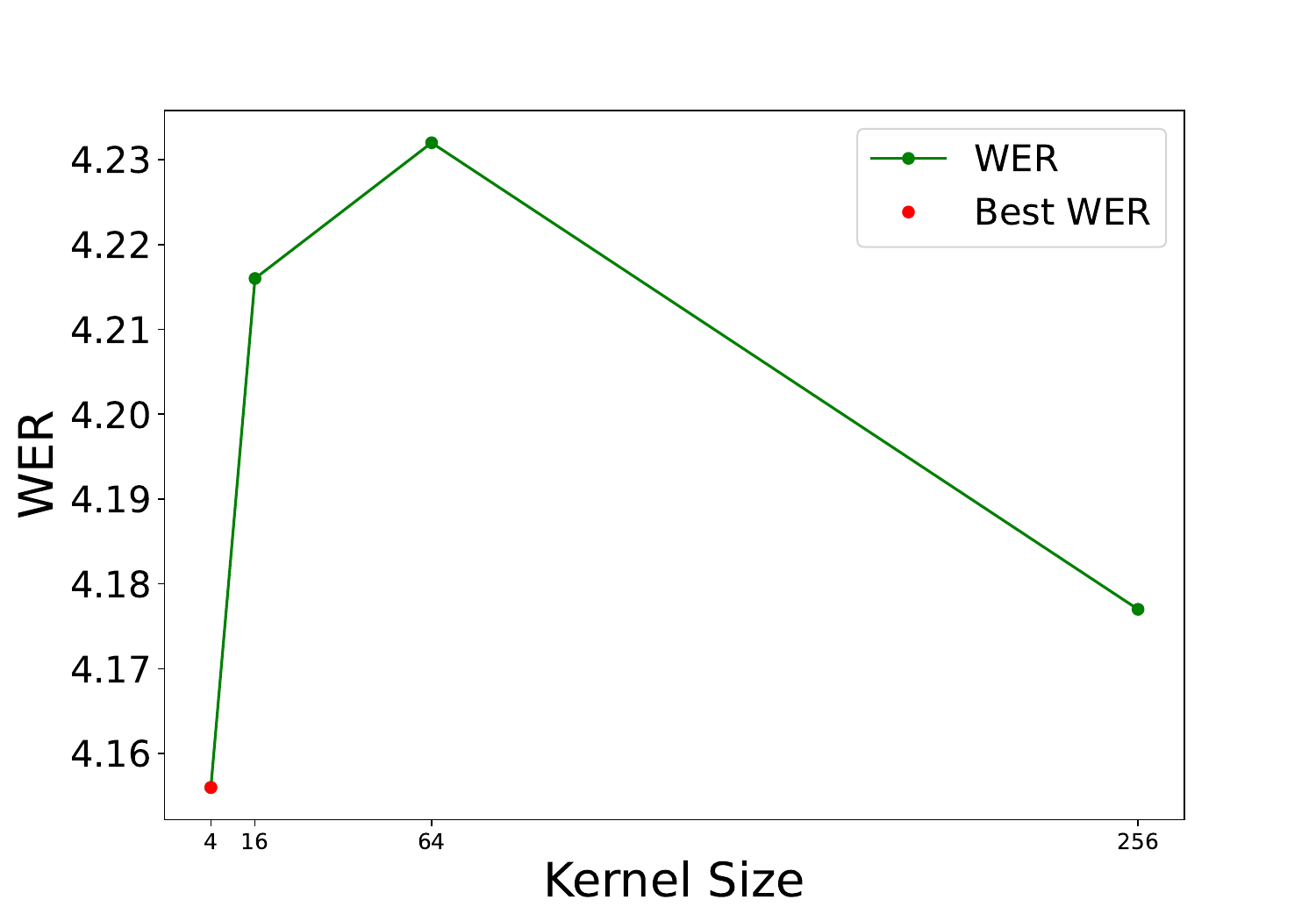}
  \vspace{-10pt}
  \caption{Word Error Rate (WER) on Librispeech dev-clean: Robustness of Our Model with Different Kernel Sizes for 1h Labeled Data.}
  \label{fig:f3}
  \vspace{-1em}
\end{figure}

This section analyzes kernel size impact on Word Error Rate (WER) using the Librispeech dev-clean dataset, particularly in the Frame stage's Echo-Transf module of Our Model(Base). Figure \ref{fig:f3} shows varying performance across kernel sizes. Notably, sizes 4 and 256 achieve lower WERs (optimal at 4.156\%), while intermediate sizes like 64 and 16 have slightly higher WERs (4.232\% and 4.216\%, respectively). This suggests a non-linear relationship between kernel size and performance.

The analysis reveals subtle WER differences among kernel sizes, implying our model's robustness. Additionally, it highlights the importance of each training stage's unique impact on performance.

\vspace{-0.5em}
\section{CONCLUSION}
\label{sec:page}

In this work, we introduce a novel variable-length attention mechanism coupled with a dynamic gating mechanism, designed to augment existing pre-trained models for enhanced Automatic Speech Recognition (ASR) performance. This enhancement is evidenced by experiments on the Librispeech corpus using 100 hours of clean training data. Our approach yields a Word Error Rate Reduction (WERR) of up to 7.7\% for Base models and 5.7\% for Large models, demonstrating robustness and parameter stability even with kernel size fine-tuning. Future research aims to explore local information utilization for further optimization and to validate the modules' effectiveness on more extensive datasets.

\bibliographystyle{IEEEtran}
\bibliography{citation}

\end{document}